\documentclass[aps,prl,floatfix,superscriptaddress,twocolumn,footinbib]{revtex4-1}
\usepackage{graphicx}
\usepackage{dcolumn}
\usepackage{bm}
\usepackage{makecell}
\usepackage{color}
\usepackage{amsmath}
\usepackage{latexsym}
\usepackage{amssymb}
\usepackage{graphics,epstopdf}
\usepackage{hyperref}
\usepackage{xcolor}
\hypersetup{
    colorlinks = true,
    linkcolor =blue,
	citecolor=blue,
	urlcolor=blue
}

\newcommand{\ed}{\end{document}}
\newcommand{\beq}{\begin{equation}}
\newcommand{\eeq}{\end{equation}}

\usepackage{mathtools}
\usepackage{comment}

\begin{document}

\title{Theory of Chirality-Induced Spin Selectivity in Trefoil-knot Molecules}

\author{Xi Sun}
 \affiliation{School of Physics and Wuhan National High Magnetic Field Center,
Huazhong University of Science and Technology, Wuhan 430074, People's Republic of China.}
\author{Shu-Zheng Zhou}
 \affiliation{School of Physics and Wuhan National High Magnetic Field Center,
Huazhong University of Science and Technology, Wuhan 430074, People's Republic of China.}
\author{Kai-Yuan Zhang}
 \affiliation{School of Physics and Wuhan National High Magnetic Field Center,
Huazhong University of Science and Technology, Wuhan 430074, People's Republic of China.}
\author{Hua-Hua Fu}
\altaffiliation{Corresponding author.\\ hhfu@hust.edu.cn}
\affiliation{School of Physics and Wuhan National High Magnetic Field Center,
Huazhong University of Science and Technology, Wuhan 430074, People's Republic of China.}
\affiliation{Institute for Quantum Science and Engineering, Huazhong University of Science and Technology, Wuhan, Hubei 430074, China.}

\date{\today}

\begin{abstract}

The origin of chirality-induced spin selectivity (CISS) remains elusive, and ultrahigh spin polarization (SP) in topologically knotted molecules is unexplained. We develop a discrete geometric spin-orbit coupling (SOC) framework for molecular junctions, resolving site-specific curvature and current-partition effects beyond continuous models. For trefoil-knot molecules, it quantifies site-resolved geometric SOC, reaching $\sim$120 meV, nearly two orders of magnitude larger than intrinsic SOC of light atoms. The substrate-coupled benzene unit, carrying the largest current and remarkable geometric SOC, dominates the CISS effect. The framework quantitatively reproduces measured SP, temperature-dependent magnetoresistance (MR) and $\Delta$MR trends, establishing discrete geometric SOC as a predictive tool for CISS in topologically knotted molecules.


\end{abstract}
\maketitle

The chirality-induced spin selectivity (CISS) effect produces robust spin polarization (SP) when electrons traverse chiral systems at room temperature and without an external magnetic field \cite{1,2,3,4,7}. Since its discovery in photoexcited electron transport through double-stranded DNA \cite{1,2}, CISS has been observed in proteins, helicenes, chiral crystals, chiral perovskites, and self-assembled systems \cite{9,10,12,13,14,17,18,19,20,23,25}, with impact extending into chemistry, energy science, and quantum information \cite{32,33,34,35,36}. Despite this progress, the microscopic origin of CISS remains under active debate \cite{37,38,39,40}.

Several mechanisms have been proposed, each focusing on different physical ingredients, including chiral-induced spin-orbit coupling (SOC), polaron transport, spinterface effects, electron-vibration coupling, and orbital polarization \cite{41,43,44,45,46,47,48,49,50,51,52,53}. These mechanisms have provided important insights into CISS in various helical systems. Recently, CISS has also been reported in systems with central, axial, and planar chirality \cite{56,57,58,63}, and even in complex supramolecular self-assemblies \cite{59,60}. In particular, experiments on trefoil-knot molecules with topologically twisted geometries have revealed CISS with ultrahigh SP \cite{61}; these molecules may be discussed in terms of topological chirality (Fig.~1(a)). Such systems go beyond the simple helical-chain picture, and involve branched, asymmetric transport pathways, entangled crossings, and electrode-coupled local geometries. A formulation that incorporates actual molecular geometry, bond angles, and branch-current partitions is therefore required.

Previous studies show that electrons moving along curved or twisted paths acquire a geometric SOC. This effect can be described within inertial or curved-space frameworks and offers a competitive explanation for CISS \cite{66,67,68,69}. However, continuous geometric-SOC descriptions do not resolve its distribution over transport sites in a realistic molecular junction, nor its combination with branch currents to govern SP. Although intrinsic and Rashba SOC are routinely included in first-principles calculations, geometric SOC arising from the electron’s curved path is not commonly treated, and capturing it requires a site-resolved description of transport geometry. Here we develop a discrete geometric SOC framework for realistic molecular junctions. Using atomic coordinates of trefoil-knot molecules \cite{61}, we quantify geometric SOC at every transport site, resolving its strength, sign, and distribution. Geometric SOC reaches $\sim$120 meV, nearly two orders of magnitude larger than the intrinsic SOC of light atoms. Combining this site-resolved SOC with a branch-current partition analysis, we identify which structural units dominate the SP. In particular, the benzene unit linked to the tip/substrate, carrying the largest current and most asymmetric geometric SOC, is key to the ultrahigh SP. This discrete framework resolves effects beyond continuous geometric-SOC models \cite{62}, including cancellation at symmetric branch points and SOC reconstruction upon electrode coupling.

Using an explicit contact geometry constrained by experiment, we obtain SP up to -72.0$\%$, close to the measured -79.5$\%$, and reproduce the temperature-dependent magnetoresistance (MR) and $\Delta$MR trends \cite{61}. These results establish the discrete geometric SOC framework as a predictive tool for CISS in trefoil-knot molecules. More broadly, this work provides a general methodology for quantifying geometric SOC in any molecular architecture with non-trivial transport geometry, and suggests design principles for chiral spintronic devices: direct current through units with large geometric SOC asymmetry, exploit electrode coupling to reconstruct SOC, and avoid symmetric branch points that cancel SOC.

{\color{blue}\emph{Discrete geometric SOC and device Hamiltonian.}}
We model tunnel junctions of $\Lambda$- and $\Delta$-type trefoil-knot enantiomers (Fig.~1(a)) using their actual atomic coordinates, with real molecular structures provided in Fig.~S1 of the Supplemental Material (SM) \cite{70}. The three lowest atoms couple to the substrate, and the topmost atom to the AFM tip. The molecular geometry defines the electron transport paths. Along a curved trajectory, the electron's direction of motion changes continuously, producing a relativistic effect in the non-inertial frame that constrains the spin orientation and produces local SP (upper inset of Fig.~1(b)). This effect manifests as a geometric SOC, governed by the local curvature of the trajectory. In the Frenet-Serret frame, the local inertial acceleration at site $i$ is $\mathbf{a}_{i}$=$-\kappa_{i} c^{2} \mathbf{N}_{i}$, with $\kappa_{i}$ the local curvature and $\mathbf{N}_{i}$ the normal vector pointing toward the center of curvature. The geometric SOC Hamiltonian is
\begin{equation}
\mathcal{H}_{\mathrm{geo}} = \sum_{i} \frac{\hbar}{4 m c^{2}} (\mathbf{a}_{i} \times \mathbf{p}_{i}) \cdot \boldsymbol{\sigma},
\end{equation}
with $\mathbf{p}_{i}$ the momentum operator, $\sigma$ the Pauli-matrices vector, $m$ the electron mass, and $c$ the light speed.

In a real trefoil-knot molecule, local transport paths can be classified into two types (Fig~1(b)): a single-channel folded path through site $i$ with bond angle $\alpha \neq \pi$ (path a: \(j \to i \to j'\)), and a branched path where three bonds intersect at site $i$ with angles $\alpha$ and $\beta$ (path b: \(j \to i \to j'_{1(2)}\)). For these local single-channel and branched paths, the tight-binding discretization gives
\begin{equation}
\mathcal{H}_{\mathrm{geo}} = i \frac{\lambda}{2} \sum_{i} \sum_{j \in (i)} f(n_{i}) \,  \sum_{j \in (i)}^{j \neq j'} c_{j}^{\dagger} \frac{\kappa_{jij'} \boldsymbol{\sigma}_{jij'} \cdot \mathbf{B}_{jij'}}{l_{jij'}} c_{i} + \mathrm{H.c.},
\end{equation}
where $\lambda = \hbar^{2}/(4m) \simeq 1.9~\mathrm{eV}\,\mathrm{\AA}^{2}$; $f(n_{i}) = 1/(n_{i} - 1)$ normalizes the branch-path contributions at site $i$ with $n_{i} \geq 2$ and $n_i$ the related coordination number; $c_j^\dagger=(c_{j,\uparrow}^\dagger,c_{j,\downarrow}^\dagger)$ is the electron creation operator at site $j$; $\kappa_{jij'}$ and $\mathbf{B}_{jij'}$ are the local curvature and binormal unit vector of the path a; $l_{j i j'} = l_{j i} + l_{i j'}$; and \(j, j' \in (i)\) denote the two endpoints of a path passing through site \(i\). For the single-channel case, the effective curvature is $\kappa_{jij'}$=$\frac{4\cos(\alpha/2)}{l_{jij'}}$. All bond lengths and angles are taken from actual molecular structures, optimized to stable geometries by first-principles calculations (Sec.~I of the SM \cite{70}, including the derivations of above equations). 

\begin{figure}[t]
\includegraphics[width=0.98\columnwidth]{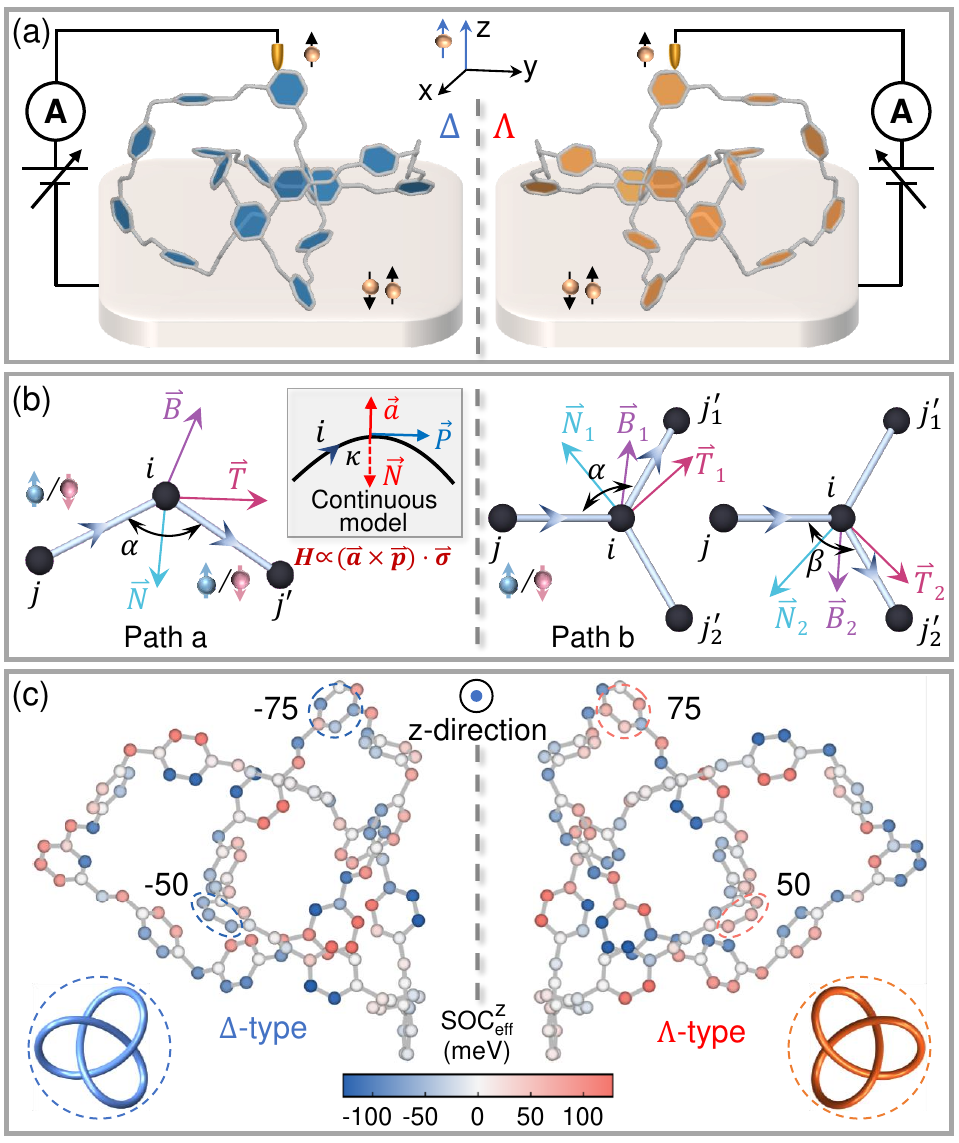}
\caption{(a) Tunnel-junction models for \(\Lambda\)- and \(\Delta\)-trefoil-knot enantiomers. Topological chirality arises from their entangled crossings. Only transport-contributing atoms are shown at actual coordinates; lowest atoms couple to the substrate, highest to an AFM tip. (b) Two transport paths: single-folded channel (Path a) and two-branched channels (Path b). \(\mathbf{N}\), \(\mathbf{B}\), \(\mathbf{T}\) are the Frenet-Serret normal, binormal, and tangent vectors; \(\mathbf{a}\) and \(\mathbf{p}\) are electron acceleration and momentum. Inset: a continuous model. (c) \(\mathrm{SOC}_{\mathrm{eff}}^{z}\) distribution at each site for both enantiomers. Red/blue denote positive/negative SOC; color intensity indicates magnitude. Insets: top-view schematics of \(\Delta\) and \(\Lambda\).}
\label{fig1}
\end{figure}

After constructing the site-resolved $\mathcal{H}_{\mathrm{geo}}$, we may build the theoretical model of the knot-molecule-based devices (Figs.~1(a)). Within the tight-binding framework, the total Hamiltonian is $\mathcal{H} = \mathcal{H}_{\mathrm{mol}} + \mathcal{H}_{\mathrm{geo}} + \mathcal{H}_{\mathrm{el}} + \mathcal{H}_{\mathrm{d}}$, where $\mathcal{H}_{\mathrm{mol}} = \sum_{i} \epsilon_{i} c_{i}^{\dagger} c_{i} + \sum_{\langle i,j \rangle} t_{ij} c_{i}^{\dagger} c_{j} + \mathrm{H.c.}$ describes intramolecular hopping, with $\epsilon_{i}$ the on-site energies set by $\epsilon_{\mathrm{C}}$=0, $\epsilon_{\mathrm{O}}$=-1.35 eV, and $\epsilon_{\mathrm{N}}$=-2.70 eV, the hopping integrals $t_{ij}$ for C-C, C-N, and C-O single and double bonds are given in Table 1 of the SM \cite{70}. These parameters are taken from established tight-binding parameterizations for conjugated organic molecules containing C, N, and O, and have been widely used to describe the electronic structure and transport properties of similar molecular system. The coupling to the AFM tip and substrate is given by $\mathcal{H}_{\mathrm{el}} = \sum_{n,\gamma} t_{n}' c_{n+1,\gamma}^{\dagger} + \sum_{i,\gamma} \Gamma_{\gamma} c_{0,\gamma}^{\dagger} c_{i} + \mathrm{H.c.}$, where $\Gamma_{\gamma}$ denotes the coupling strength and $\gamma$ labels the semi-infinite metallic leads. We take $\Gamma_{\mathrm{tip}}$=$\Gamma_{\mathrm{substrate}}$=0.1 eV, consistent with typical metal–molecule couplings. The term $\mathcal{H}_{\mathrm{d}} = i \sum_{i,k} \Gamma_{d} c_{k}^{\dagger} c_{i} + \mathrm{H.c.}$ describes the electron decoherence process of the experimental environment with $\Gamma_{d} = 0.015~\mathrm{eV}$ \cite{71,72}. Since the molecules consist of light atoms, many-body effects have little influence on the SP. We therefore neglect explicit Coulomb interactions but retain the on-site Coulomb term $\mathcal{H}_U=\sum_{i}U_{i}n_{i,\uparrow}n_{i,\downarrow}$ in supplemental calculations. Using the nonequilibrium Green's function transport theory \cite{73,74,75,76}, we calculate the local bond current $I_{ji}$ and total charge currents $I_{\uparrow}$ and $I_{\downarrow}$ for the tip magnetization parallel and antiparallel to the z axis, respectively, and obtain $\mathrm{SP} = \frac{I_{\uparrow} - I_{\downarrow}}{I_{\uparrow} + I_{\downarrow}}$ (Sec. III of the SM \cite{70}).

\begin{figure}[t]
\includegraphics[width=0.99\columnwidth]{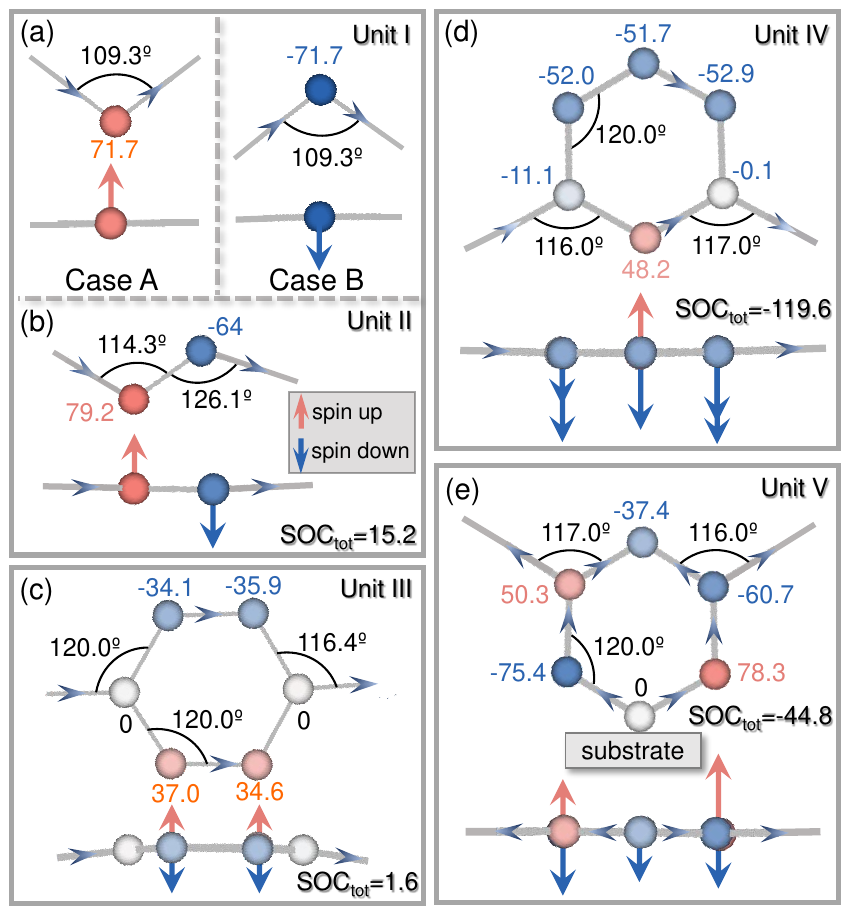}
\caption{Five transport units with lattice-resolved $\mathrm{SOC}_{\mathrm{eff}}^{z}$. The total $\mathrm{SOC}_{\mathrm{eff}}^{z}$ strength of each unit is denoted by SOC$_\mathrm{tot}$. (a) Unit I: a single-folded channel; its effective geometric SOC is perpendicular to the bond plane. (b) Unit II: a double-folded channel with opposite curvatures, giving SOC$_\mathrm{tot}$=15.2 meV. (c) Unit III: a benzene-ring-linked symmetric channel with two branch paths; the symmetric currents yields SOC$_\mathrm{tot}$=1.6 meV. (d) Unit IV: a benzene-ring-linked asymmetric channel with two branch paths; the asymmetric current yields SOC$_\mathrm{tot}$=119.6 meV. (e) Unit V: a benzene-ring-linked triple asymmetric channel with one site coupled to the substrate; SOC$_\mathrm{tot}$=-44.8 meV. Upper panels: view along the $z$-axis; lower panels: side view.}
\label{fig2}
\end{figure}

{\color{blue}\emph{Effective curvature vector and SOC distribution.}} 
Although Eq.~(2) defines the geometric SOC at each site, the actual contribution to the SP depends on how electrons traverse the local branches. Electrons do not pass through all local paths with equal probability, so the bare local curvature \(\kappa_{jij'}\) alone does not reflect the effective SOC weighted by the transport current. To relate the actual molecular geometry to CISS, we introduce a lattice-resolved effective curvature vector that incorporates the local bond currents. For the path a (Fig.~1(b)), we define the path curvature vector $\vec{c}_{a}$=$\frac{\kappa_{a}}{l_{a}} \vec{B}_{a}$, where $\kappa_{a}$=$\kappa_{jij'}$ is the local curvature, \(l_{a} = l_{ji} + l_{ij'}\) is the total path length, and \(\vec{B}_{a}\) is the binormal unit vector. The factor \(1/l_{a}\) makes curvature vectors of different path lengths directly comparable and follows naturally from the discretization of the continuum geometric SOC. This vector characterizes both the magnitude and direction of the local curvature of path a.

To weight each path by its actual contribution to transport, we use the local bond currents \(I_{ji}\) and \(I_{ij'}\). The probability that an electron traverses the path a is proportional to the product of the currents on the two bonds, so we define the path weight $W_{a} = I_{ji} I_{ij'}$. After normalization over all paths passing through site \(i\), $\omega_{a} = \frac{W_{a}}{\sum_{a \in A_{i}} W_{a}}$, where \(A_{i}\) is the set of all paths passing through site \(i\). The effective curvature vector at site \(i\) is then $\vec{c}_{i} = \sum_{a \in A_{i}} \omega_{a} \vec{c}_{a}$. This vector represents the current-weighted average of the local bending at site \(i\), including both its magnitude and direction. Projecting \(\vec{c}_{i}\) onto the tip magnetization direction \(\hat{z}\) yields the site-resolved effective SOC, $\mathrm{SOC}_{\mathrm{eff}}^{z} = \frac{\lambda}{4} \vec{c}_{i} \cdot \hat{z}$. The factor \(1/4\) arises from two sources: each path is counted twice in the sum over \(j\) and \(j'\) (forward and backward directions), and the Hamiltonian is Hermitian-symmetrized with respect to the tangential momentum and the local spin operator (detailed derivations provided in Sec.~IV of the SM~\cite{70}).


\begin{figure*}[t]
\includegraphics[width=1.95\columnwidth]{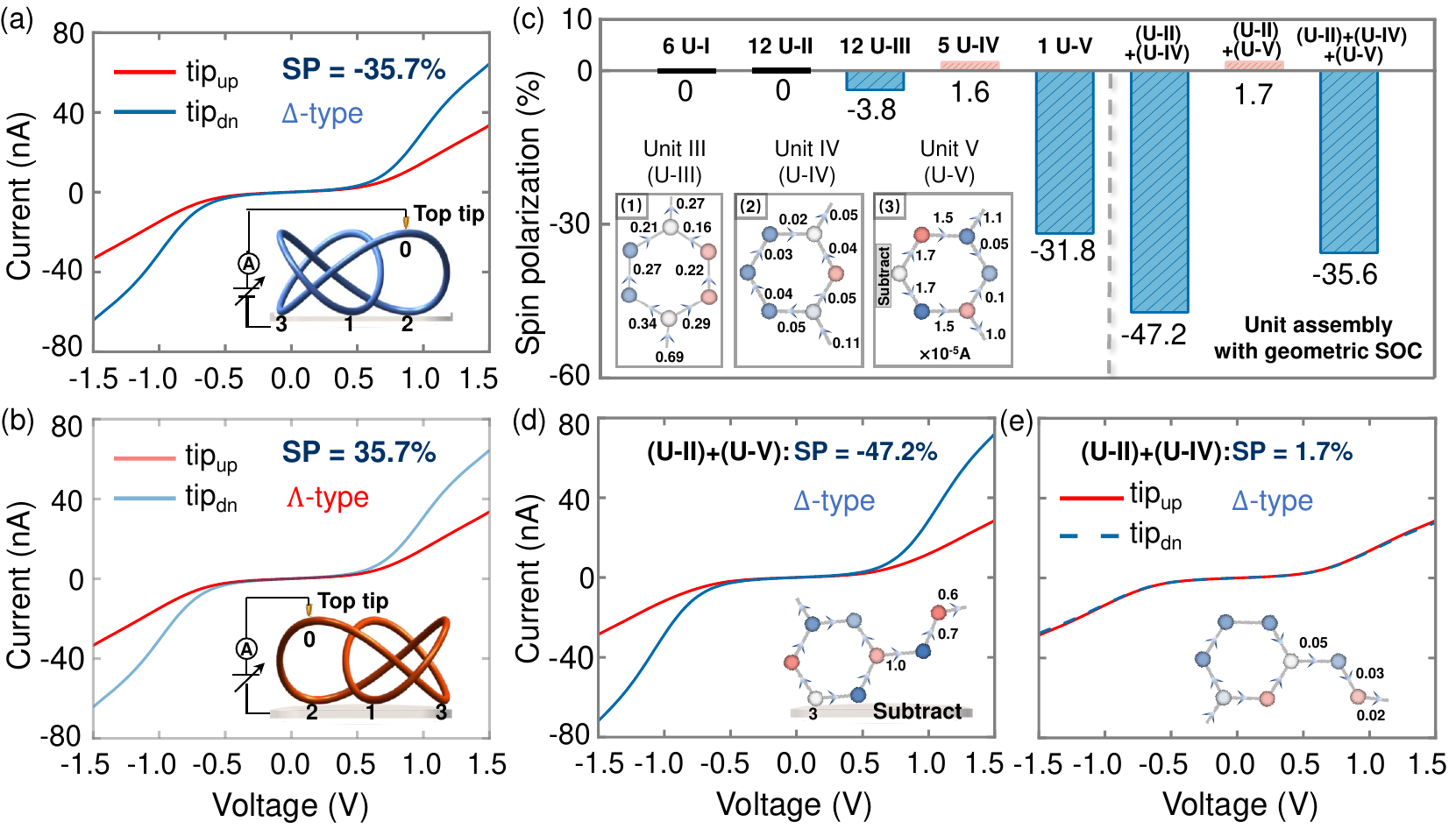}
\caption{(a), (b) Current-voltage curves of $\Delta$- and $\Lambda$-trefoil-knot molecular junctions for tip magnetization up (red) and down (blue); SP $=-35.7\%$ (a) and $35.7\%$ (b). Insets: side views. (c) SP contributions of five transport units (U-I--U-V) and three combinations in the $\Delta$-trefoil molecule. Left bars: each unit's contribution with geometric SOC assigned only to that unit; right bars: three combinations. Lower-left insets: current direction and magnitude ($\times10^{-5}$ A) in U-III--U-V. (d), (e) Current versus voltage with geometric SOC retained only for two combinations: (d) (U-II)+(U-V), SP $=-47.2\%$; (e) (U-II)+(U-IV), SP=1.7$\%$. Insets: unit combinations coupled to substrate. Related data for $\Lambda$ are provided in Fig. S2 of the SM \cite{70}.}
\label{fig3}
\end{figure*}

Figure~1(c) shows the distribution of \(\mathrm{SOC}_{\mathrm{eff}}^{z}\) over all transport sites for the \(\Delta\)- and \(\Lambda\)-type enantiomers. Three features are notable. First, \(\mathrm{SOC}_{\mathrm{eff}}^{z}\) is highly nonuniform, reaching about \(120\)~meV, nearly two orders of magnitude larger than the intrinsic SOC of light atoms, and remains sizable over multiple transport sites. Second, at corresponding sites of the two enantiomers, \(\mathrm{SOC}_{\mathrm{eff}}^{z}\) has nearly equal magnitude but opposite sign, directly linking chirality reversal to SP reversal. Third, its strength is correlated with the local curvature, while its sign is determined by the bending direction. At symmetric threefold branch points, the contributions from different branches cancel: when the three path curvature vectors \(\kappa_{a}\vec{B}_{a}\) have equal magnitude and are separated by \(120^{\circ}\), their vector sum vanishes, leaving negligible net SOC.

These results show that certain local structures consistently produce large or small net SOC. To identify which structural units dominate the SP, we classify the transport-active units into five types: unit I, a single-folded channel (Fig.~2(a)); unit II, a double-folded with opposite curvatures (Fig.~2(b)); unit III, a benzene-ring-linked symmetric channel (Fig.~2(c)); unit IV, a benzene-ring-linked asymmetric channel with two branch paths (Fig.~2(d)); and unit V, a benzene-ring-linked triple asymmetric channels with one site coupled to the substrate/tip (Fig.~2(e)). Unit I, the simplest unit, has a bond angle of \(109.3^{\circ}\). Depending on its opening direction, the $\mathrm{SOC}_{\mathrm{eff}}^{z}$ along this path is about \(+71.7\)~meV (case A) or \(-71.7\)~meV (case B), with the resulting SP perpendicular to the bond plane (lower panel of Fig.~2(a)). Unit II has two folds with similar angles (\(114.3^{\circ}\) and \(126.1^{\circ}\)) but opposite opening directions, giving nearly equal while opposite $\mathrm{SOC}_{\mathrm{eff}}^{z}$; its SP contribution is therefore weak. Unit III comprises symmetric benzene rings, and the current maintains axial symmetry; $\mathrm{SOC}_{\mathrm{eff}}^{z}$ at the four axially symmetric sites is nearly equal but opposite, while that at current entry/exit sites is negligible. Thus, units similar to unit III contribute negligibly to SP.

The connection topology markedly breaks the axial symmetry of the current through the benzene ring. Unit~IV (Fig.~2(d)), the dominant structural motif, carries current mainly through a multi-site upper channel where the $\mathrm{SOC}_{\mathrm{eff}}^{z}$ is large and of the same sign at each site, except for one lower-channel site of opposite sign. When the knot couples to substrate or tip, an additional unit~V appears: its configuration is identical to unit~IV, but one site binds to the tip/substrate, altering the current direction and distribution and reconstructing the geometric SOC within the ring (Fig.~2(e)). At the electrode-coupled site, $\mathrm{SOC}_{\mathrm{eff}}^{z}$ vanishes; although the geometric SOC is redistributed, its total magnitude remains significant, making Unit~V an important source of SP for conducting electrons.

{\color{blue}\emph{Current partition and unit contributions.}} To identify the origin of the SP in the $\Delta$- and $\Lambda$-trefoil-knot junctions (Figs.~1(a) and 1(b)), we calculate the tunneling currents for tip magnetization up and down (Figs.~3(a) and 3(b)). Insets show side views: site 0 couples to the AFM tip and sites 1-3 to the substrate. For both enantiomers, $I_{\mathrm{up}}$ and $I_{\mathrm{dn}}$ are well separated and obey Onsager reciprocity. The currents reverse between enantiomers, giving $\mathrm{SP}$=-35.7$\%$ ($\Delta$) and +35.7$\%$ ($\Lambda$). This opposite SP is tied to topological chirality, confirming CISS in knotted molecules. We then decompose the SP into contributions from individual transport units.


Geometric SOC alone does not guarantee high SP. The current direction within each unit and its partition among branching paths are equally important: a branch with strong geometric SOC contributes little if its current is weak. In Fig.~3(a), paths \(L_{3\to 0}\) and \(L_{2\to 0}\) dominate transport, whereas \(L_{3\to 1}\) and \(L_{2\to 1}\) carry weaker currents. Thus, the SP contribution of each unit is governed by the product of its geometric-SOC asymmetry and its current partition.


To identify which units dominate, we retain geometric SOC only on selected units as a theoretical decomposition. The $\Delta$-type knot contains 6 Unit I, 12 Unit II, 12 Unit III, and 5 Unit IV units, plus 1 substrate-coupled Unit V. With geometric SOC only on Unit I, the two opposite opening directions (Cases A and B) cancel exactly, giving no net SP. Unit II behaves similarly, with half opposite to the other half, yielding negligible SP; these correspond to the two zero-height bars in Fig.~3(c). For Unit III, symmetric current through the two branches produces nearly equal but opposite SOC at paired sites [Fig.~3(c), inset (1)], and slight current asymmetry leaves only $-3.8\%$ SP. For Unit IV, the two asymmetric paths have SOC of $-52.0$ and $48.2$~meV, but the left path has weaker current, so the two contributions nearly cancel, giving only $1.6\%$ SP. Thus, Units I--IV are not the key units for high SP.


The paths \(L_{3\to 0}\) and \(L_{2\to 0}\) dominate transport and both include the substrate-coupled benzene unit (Unit V). This unit has large geometric SOC and is directly bonded to the substrate, so both benzene-ring branches carry substantial currents [Fig.~3(c), inset (3)]. With SOC only on Unit V, SP reaches \(-31.8\%\) [fifth bar, Fig.~3(c)]. Adding neighboring Unit II increases it to \(-47.2\%\) [Fig.~3(d)]: Unit II alone gives no SP, but redistributes current through Unit V, enhancing its SOC asymmetry. In contrast, the non-substrate-coupled Unit IV with Unit II yields only \(1.7\%\) [Fig.~3(e)]; adding Unit V raises it to \(-35.6\%\) [Fig.~3(c)]. Thus, the substrate-coupled Unit V governs the high SP.


\begin{figure}[t]
\includegraphics[width=0.99\columnwidth]{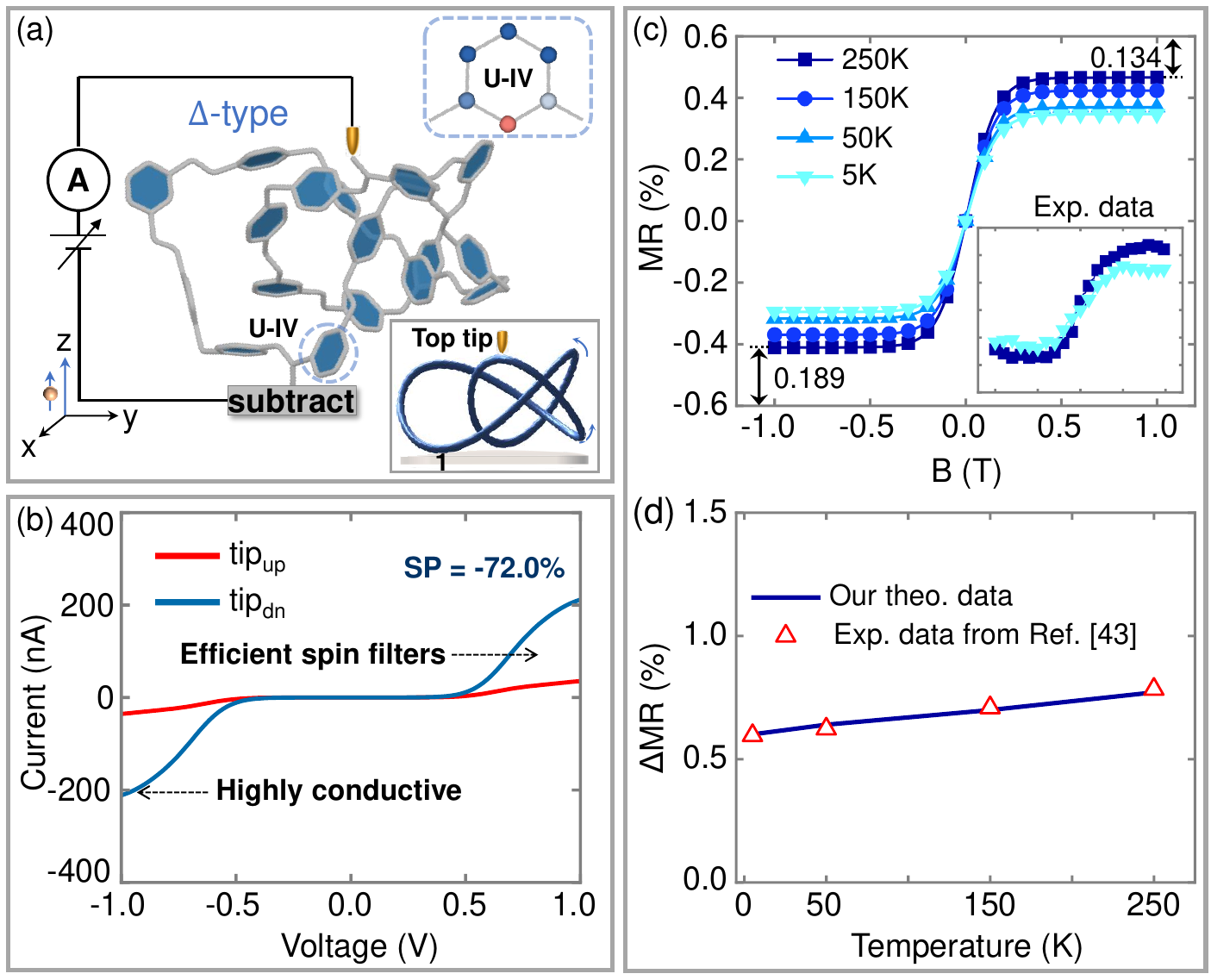}
\caption{(a) Optimized $\Delta$-type trefoil-knot molecular junction. The molecule couples to the substrate via one lowest atomic site and to the tip via one terminal site; both coupling sites lie as close as possible to U-IV units. (b) Current-voltage curves for up and down magnetization for the optimized device. The SP reaches -$72.0\%$, indicating excellent spin filtering. Inset: experimental data for a $\Delta$-type trefoil-knot molecule, with SP -$79.5\%$. (c) MR versus magnetic field B at 5, 50, 150, and 250 K. Inset: experimental MR for the same type of trefoil-knot molecule at 5 and 250 K. (d) Calculated and measured $\Delta$MR versus temperature. All experimental data are from Ref.~\cite{61}.}
\label{fig4}
\end{figure}

Previous experiments attributed the electrode dependence of CISS to the metal's high SOC or spinterface coupling \cite{49,50,77,78}. Our calculations reveal an additional mechanism: current partitioning at the molecule-electrode contact, combined with the geometric SOC of the coupled unit, governs the high SP. These mechanisms are complementary and together may account for the experimental observations.


{\color{blue}\emph{Device optimization and experimental comparison.}} Guided by the current-partition analysis, we consider an optimized geometry in which the \(\Delta\)-type trefoil-knot molecule couples to the substrate through a single lowest atomic site, with both contacts placed near the U-IV units (Fig.~4(a)). This two-port design restricts the substrate-to-tip current to two twisted channels, avoiding current spreading and directing it through units with strong geometric SOC asymmetry. The optimized geometry is not a fit to the experimental SP; it follows from the requirement that the dominant transport paths include the substrate-coupled Unit V.

This geometry is consistent with the experimental conditions of Ref~\cite{61}, where the film thickness (\(\sim 3.0\)~nm) is close to the molecular length and mCP-AFM measurements suggest anchoring through a few lowest atomic sites, with the molecular body tilted toward the tip rather than lying flat with all three lowest sites coupled. Figure~4(b) shows the current-voltage curves for tip magnetization up and down. The SP reaches -72.0$\%$, close to the measured -79.5$\%$, indicating efficient spin filtering. At $V$=1.0 V the current reaches 200 nA, corresponding to a high conductance. The remaining difference of about 7.5$\%$ may arise from many-body effects, conformational disorder, or details of the electrode coupling not captured by the single-particle model. Figure~4(c) shows the calculated MR versus magnetic field at different temperatures (Sec. V of the SM \cite{70}). The MR exhibits an asymmetric field response, $\mathrm{MR}(B)$$\neq$$\mathrm{-MR}(-B)$, which originates from the interplay between the chiral geometry and the applied field. The calculated \(\Delta\mathrm{MR}\), defined as $\Delta\mathrm{MR}(\%)$=$|\mathrm{MR}(\%)|_{-1.0\,\mathrm{T}}$+$|\mathrm{MR}(\%)|_{+1.0\,\mathrm{T}}$, agrees well with the experimental data (Fig.~4(d)). This agreement supports the geometric-SOC mechanism and highlights the role of current partitioning at the electrode contact. The optimization strategy developed here, restricting current to channels with strong geometric SOC asymmetry, is general and can be applied to other chiral molecular junctions. It suggests two design principles: anchor the molecule through a single contact to avoid current spreading, and place the contacts near asymmetric units that carry large geometric SOC.

In summary, we have developed a discrete geometric SOC framework for CISS in realistic trefoil-knot molecular junctions. The framework incorporates the actual transport pathways and branch-current partitions, providing a site-resolved geometric SOC and its relation to the SP. Unit-resolved analysis shows that the SP is governed jointly by the geometric SOC asymmetry and the current partition through each unit, with the substrate-coupled benzene unit dominating the high SP. The calculated SP and temperature-dependent MR and $\Delta$MR trends are well consistent with experimental observations. These results provide strong evidence for the geometric-SOC mechanism of CISS in topologically chiral knots and suggest two design principles for chiral molecular spintronics: anchoring through a single contact to avoid current spreading, and placing contacts near asymmetric units with large geometric SOC.


{\color{blue}\emph{Acknowledgments.}} This work is supported by the authors' personal resource.

\end{document}